# X-ray Anomalous Scattering Study of a Charge-Ordered State in $NaV_2O_5$

H. Nakao,[1,*] K. Ohwada,[1] N. Takesue,[1] Y. Fujii,[1] M. Isobe,[2] Y. Ueda,[2]
M. v. Zimmermann,[3] J. P. Hill,[3] D. Gibbs,[3] J. C. Woicik,[4] I. Koyama,[5] and Y. Murakami[5]

[1] *Neutron Scattering Laboratory, Institute for Solid State Physics, The University of Tokyo, 106-1 Shirakata, Tokai, Ibaraki 319-1106, Japan*

[2] *Material Design and Characterization Laboratory, Institute for Solid State Physics, The University of Tokyo, 7-22-1 Roppongi, Minato-ku, Tokyo 106-8666, Japan*

[3] *Department of Physics, Brookhaven National Laboratory, Upton, NY 11973-5000*

[4] *National Institute of Standards and Technology Gaithersburg, MD 20899*

[5] *Photon Factory, Institute of Materials Structure Science, High Energy Accelerator Research Organization (KEK), Tsukuba, 305-0801, Japan*

Charge ordering of $V^{4+}$ and $V^{5+}$ in $NaV_2O_5$ has been studied by an X-ray diffraction technique using anomalous scattering near a vanadium $K$-absorption edge to critically enhance a contrast between the two ions. A dramatic energy dependence of the superlattice intensities is observed below $T_C = 35$ K. Consequently, the charge ordering pattern is the fully-charged *zigzag*-type ladders with the unit cell $2a \times 2b \times 4c$, but not the *chain*-type originally proposed for the spin-Peierls state. Charge disproportionation suggested in our model as the average valence $V^{4.5 \pm \delta_c/2}$ is observed below $T_C$, showing continuous variation of $\delta_c$ as a function of temperature.

The inorganic crystal $\alpha'$-$NaV_2O_5$ has attracted much attention as a quasi-1D spin-$\frac{1}{2}$ Heisenberg antiferromagnet (AF) system, which exhibits a steep decrease of its magnetic susceptibility below $T_C = 35$ K suggesting a spin-Peierls (SP) transition.[1] According to the original structure analysis,[2] the space group was a noncentrosymmetric $P2_1mn$, with magnetic chains of $V^{4+}$ ($s = \frac{1}{2}$), separated by nonmagnetic chains of $V^{5+}$ ($s = 0$) running along the $b$-axis. The magnetic chains were playing a role as those of the Heisenberg AF model above $T_C$ to cause the SP phase transition at $T_C$. Moreover, the observation of atomic displacements with the modulation wave vector $\boldsymbol{q} = (\frac{1}{2}, \frac{1}{2}, \frac{1}{4})$ by X-ray and spin gap formation at the $(1, \frac{1}{2}, 0)$ by neutron scattering strongly suggested the SP state below $T_C$.[3] NMR[4] and Raman scattering[5] experiments also supported this transition.

In 1998, the structural refinement at room temperature ($> T_C$) revealed the space group as $Pmmn$. This symmetry allows only one kind of V sites where the ions have an average valence of $V^{4.5+}$,[6,7] and they form spin ladders oriented along the $b$-axis, resulting in the so-called quarter filled ladder. On the basis of this new structure, the V-O-V orbital which forms a rung along the $a$-axis has only one $d$-electron, so that the insulating state is retained, and the chains along the $b$-axis behave as the Heisenberg AF above $T_C$.[6,8] The NMR experiment indicated existence of two inequivalent V sites of $V^{4+}$ and $V^{5+}$ below $T_C$, since the signals merge into the one signal of $V^{4.5+}$ sites above $T_C$.[9] This phase transition is no longer an ordinary SP transition but is a spin-charge-lattice-coupled system, which is of a novel case. Lüdecke *et al.*[10] and Boer *et al.*[11] have recently succeeded in analysis of the structure below $T_C$ as the space group $Fmm2$. In their analysis, however, the charge ordered state of the V ions was not directly determined.

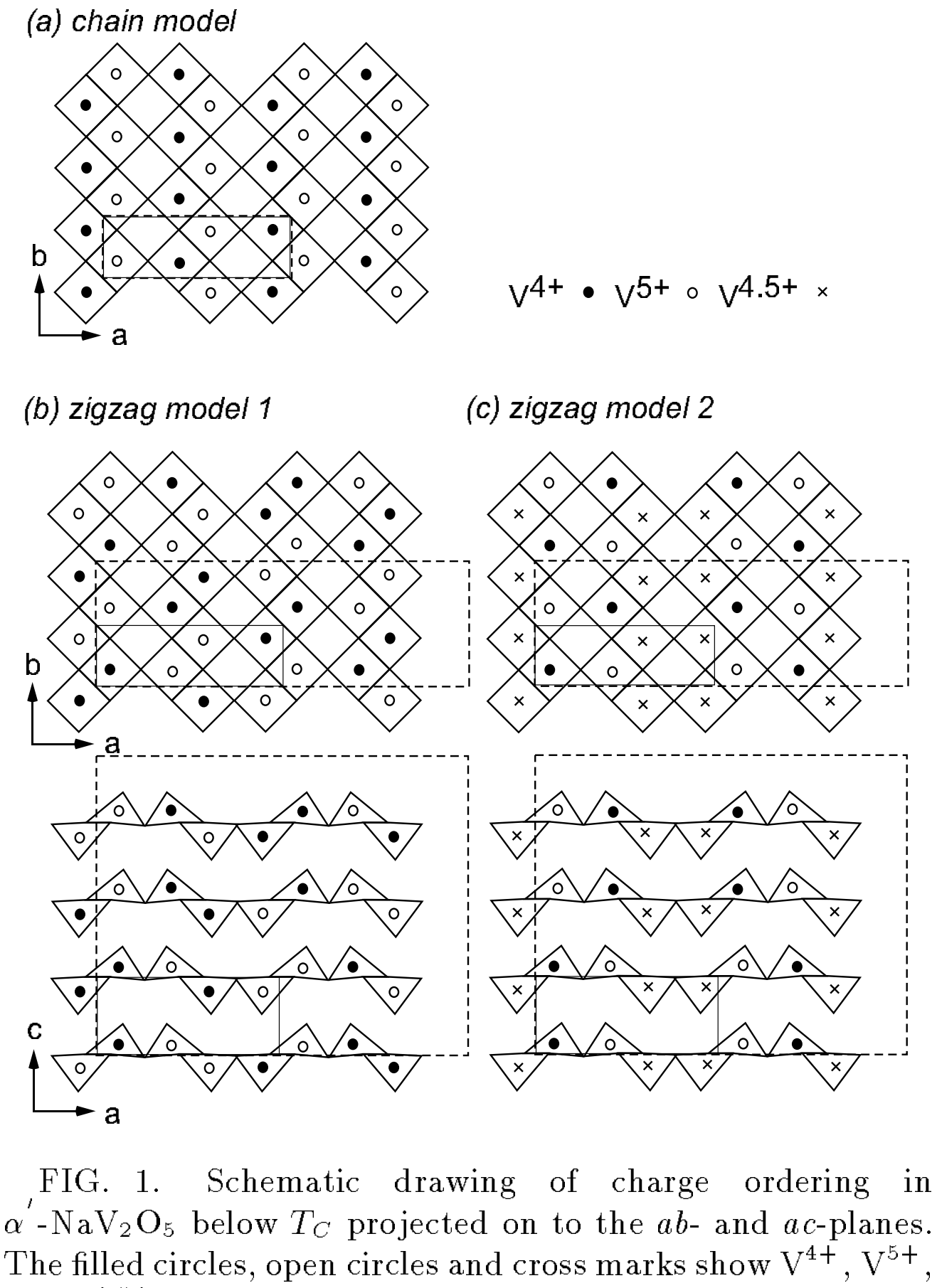

FIG. 1. Schematic drawing of charge ordering in $\alpha'$-$NaV_2O_5$ below $T_C$ projected on to the $ab$- and $ac$-planes. The filled circles, open circles and cross marks show $V^{4+}$, $V^{5+}$, and $V^{4.5+}$ sites, respectively. The solid and broken rectangles indicate the unit cell at room temperature and low temperature phase, respectively. (a) *chain model*, (b) *zigzag model 1*, and (c) *zigzag model 2*.

Two models of the charge order have been theoretically proposed; the one is based on the 1D $V^{4+}$ chains along the $b$-axis (*chain model* shown in Fig. 1 (a)), and the other is characterized as the *zigzag* pattern formed by $V^{4+}$ ions along the ladders (*zigzag model 1* shown in Fig. 1 (b)). Based on these two models, the interplay between the charge ordering and the spin-gap formation was studied, indicating that the charge degree of freedom plays an important role in $NaV_2O_5$.[12–14]

In this letter, we study the structure of charge ordering in $NaV_2O_5$ by using X-ray anomalous scattering near the vanadium $K$-absorption edge ($E_A \sim 5.465$ keV) to preferentially probe the V ions. Dramatic energy dependence of superlattice reflections with wave vector $\boldsymbol{q} = (\frac{1}{2}, \frac{1}{2}, \frac{1}{4})$ is observed, and hence the unit cell relevant to charge order is $2a \times 2b \times 4c$, which rules out the *chain* model ($a \times b \times c$). Therefore, we consider the following two models to explain our experimental data; one assumes a fully charge ordered state as shown in Fig. 1(b) (*zigzag model 1*), the other has a charge disordered ladder to satisfy the space group $Fmm2$ as shown in Fig. 1(c) (*zigzag model 2*); The former is based on the NMR results[9] and inconsistent with the space group $Fmm2$,[10] and the latter involves the three kinds of V ions, $V^{4+}$, $V^{5+}$, and $V^{4.5+}$. The model calculation of the fully-charged *zigzag model 1* can reproduce our results better. The order parameter of the charge ordered state $\delta_c$ is also determined as a function of temperature.

High-quality $NaV_2O_5$ single crystals were grown by a flux method.[15] A typical crystal is $1 \times 5 \times 0.6$ mm$^3$ ($a \times b \times c$) in size. The crystal has an orthorhombic structure with the lattice constants $a = 11.3$ Å, $b = 3.61$ Å and $c = 4.80$ Å at room temperature. The experiments were carried out both at the National Synchrotron Light Source (NSLS) of Brookhaven National Laboratory and at the Photon Factory (PF) of KEK. The absorption spectrum was measured on the beam line X23A2(NSLS) with a Si(311) monochromator. The X-ray scattering experiments were performed at X22C(NSLS) with a Ge(111) monochromator and at multipole-wiggler beam line 16A2 (PF) with a Si(111) monochromator. A four-circle diffractometer equipped with a closed-cycle helium refrigerator was used.

In order to determine the charge ordered structure in $NaV_2O_5$, anomalous scattering techniques were applied to enhance the slight difference of atomic scattering factors between $V^{4+}$ and $V^{5+}$. The incident energy dependence of atomic scattering factor is generally described by $f(E) = f_0 + f^{'}(E) + if^{''}(E)$, where $f_0$ is the Thomson scattering factor. The $f^{'}$ and $f^{''}$ are real and imaginary parts of the anomalous scattering factor, respectively. The $f^{''}$ can be obtained directly from an absorption spectrum $\mu(E)$ while the $f^{'}$ is obtained by the Kramers-Krönig transformation of the $f^{''}$.[16] Fortunately, $CaV_2O_5$ and $V_2O_5$ have the same crystal structure as $NaV_2O_5$ and contain only $V^{4+}$ and only $V^{5+}$ ions, respectively. Their spectra are shown in Fig. 2 (a) and (b). The absorption edge energy($E_A$) of two ions differ by about 1.8 eV due to a chemical shift. A peak in $f^{''}$ at 5.47 keV is a pre-edge feature corresponding to the $1s \rightarrow 3d$ transition of the V ion, consistent with a previous report.[17] Such a strong anomalous scattering intensity is influenced by the lack of local inversion symmetry on the vanadium sites. This state allows hybridization of vanadium $4p$ with the $3d$ states, leading to a significant dipole transition at the energy.[18] As indicated in Fig. 2 (c), a difference between the two spectra is small at $E \ll E_A$, while it is critically enhanced with a large modulation near $E_A$. Therefore, the difference of one electron between the $V^{4+}$ and $V^{5+}$ sites should be easily detected from the scattering intensity as a function of the energy across $E_A$.

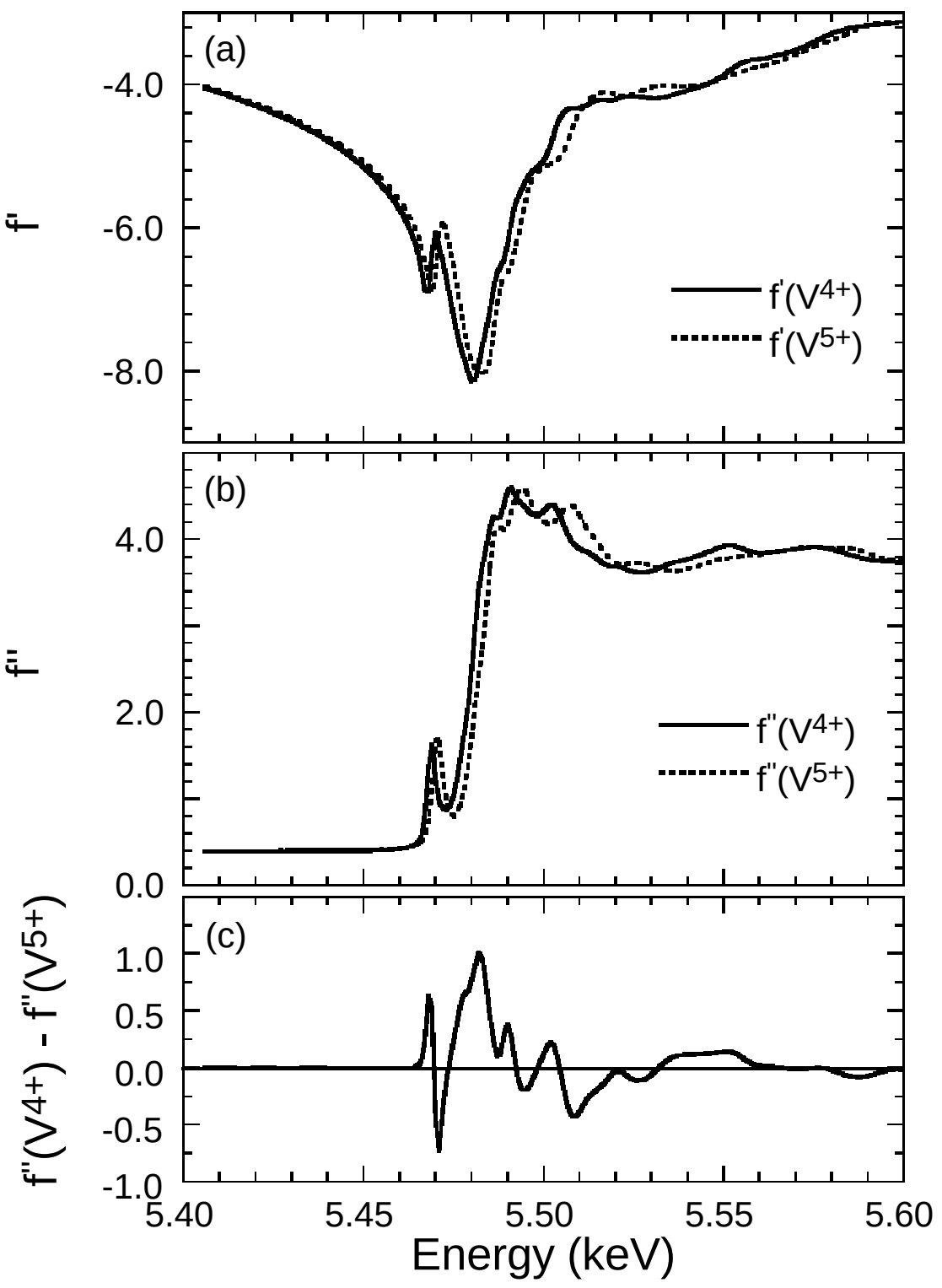

FIG. 2. (a), (b) The energy dependence of anomalous scattering factors $f^{'}$ and $f^{''}$ for $V^{4+}$ (solid line) and $V^{5+}$ (broken line), respectively. (c) The energy dependence of the difference $f^{''}(V^{4+}) - f^{''}(V^{5+})$.

The energy dependence of $(\frac{h}{2}, \frac{k}{2}, \frac{\ell}{4})$-type structure factor is calculated as follows:

$$F(\tfrac{h}{2}, \tfrac{k}{2}, \tfrac{\ell}{4}) \propto i(\boldsymbol{\delta} \cdot \boldsymbol{Q})\Big\{a(\boldsymbol{Q})\Big[f(V^{4+}) + f(V^{5+})\Big] + b(\boldsymbol{Q})\Big[f(V^{4+}) - f(V^{5+})\Big]\Big\} + c(\boldsymbol{Q})\delta_c\Big[f(V^{4+}) - f(V^{5+})\Big] + D(\boldsymbol{Q}). \quad (1)$$

The first three brackets are directly related to the V ions. $a(\boldsymbol{Q})$, $b(\boldsymbol{Q})$, and $c(\boldsymbol{Q})$ are phase factors related to V atomic positions in the unit cell. The $f$ indicates the atomic scattering factors of $V^{4+}$ and $V^{5+}$. $D(\boldsymbol{Q})$ is a structure factor related to the other ions which are approximately independent of $E$ near the V edges. $\boldsymbol{\delta}$ is an atomic displacement of the V ions for $T < T_C$. The first two brackets are multiplied by $\delta$, so called "$\delta$ term". $\delta_c$ is the order parameter of charge ordering so that $\delta_c = 1$ corresponds to full charge disproportionation. The third bracket arising from the charge disproportionation is called the "$\delta_c$ term". Nonzero $\delta_c$ term making the intensity contribution of $f(V^{4+}) - f(V^{5+})$ is crucial to determine the charge ordering pattern. The difference term is also included in the $\delta$ term. However, its contribution is negligible since the term is multiplied by the atomic displacement $\delta$, and the first bracket including $f(V^{4+}) + f(V^{5+})$ is $10^{1\sim2}$ times larger. As a result, any intensity from the $\delta_c$ term near $E_A$, which is comparable to the $\delta$ term, should provide direct information about charge ordering.

The intensities observed at $(\frac{15}{2}, \frac{1}{2}, \frac{1}{4})$ and $(\frac{13}{2}, \frac{1}{2}, \frac{1}{4})$ as a function of X-ray energy are given by the dots in Fig. 3. A large intensity modulation has been observed at $(\frac{15}{2}, \frac{1}{2}, \frac{1}{4})$ near $E_A$. This result evidences the dominant contribution from the $\delta_c$ term, indicating that the unit cell size for charge order is $2a \times 2b \times 4c$. On the other hand, the intensity at $(\frac{13}{2}, \frac{1}{2}, \frac{1}{4})$ is not enhanced near $E_A$, so that the intensity is dominated by the $\delta$ term. The charge ordering pattern below $T_C$ is discussed in the following paragraph on the basis of these results.

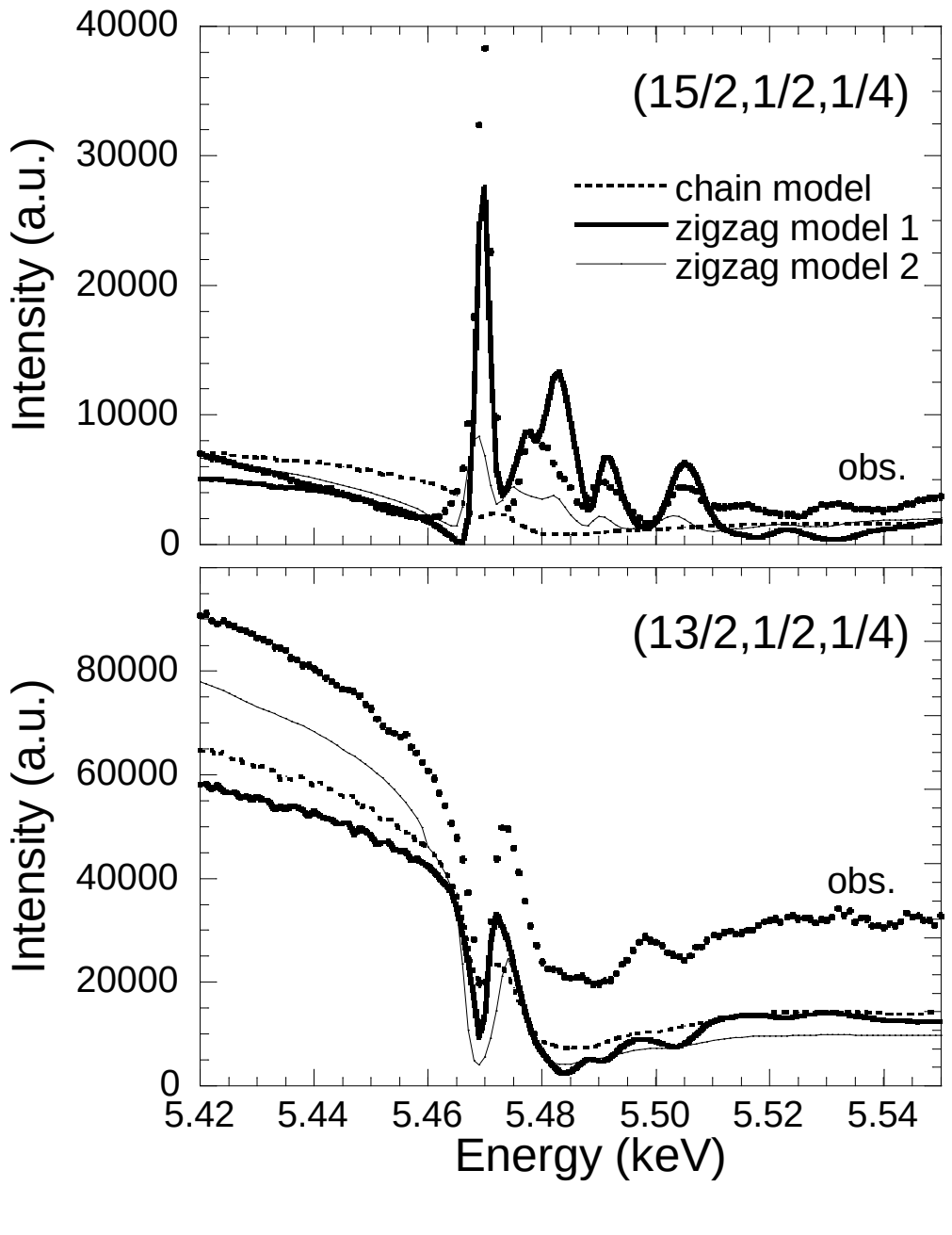

FIG. 3. Energy dependence of the superlattice reflections $(\frac{15}{2}, \frac{1}{2}, \frac{1}{4})$ and $(\frac{13}{2}, \frac{1}{2}, \frac{1}{4})$ near $E_A$ observed at $T = 8$ K. Solid, thin solid, and dotted curves are the calculations based on the *zigzag model 1, 2* and *chain model*, respectively.

Considering the results of the NMR[9,19] and X-ray structure analysis[10] below $T_C$, the space group $Fmm2$ allows three kinds of V ions in the $ab$ plane, one $16e$ site and two $8d$ sites. If $V^{4+}$ ($V^{5+}$) occupies $16e$ site, both $8d$ sites must be occupied by $V^{5+}$ ($V^{4+}$). As a result, the $V^{4+}$ and $V^{5+}$ ladders are formed in the $ab$ plane. However, such charge ordering does not satisfy doubling of the period along the $b$-axis observed experimentally. In fact, this model calculation does not reproduce the nonzero $\delta_c$ term at $(\frac{15}{2}, \frac{1}{2}, \frac{1}{4})$. Therefore we consider two other models. The first model is based on the NMR result with a fully charge ordered state, i.e. $\delta_c = 1$. As the structural restriction, we also use the extinction rule and the rule of charge ordering; one V ion on a rung is $V^{4+}$ ($V^{5+}$) and the other one is $V^{5+}$ ($V^{4+}$) because the V-O-V orbital on a rung has one electron above $T_C$ to keep the insulating state.[6,8] As a result, we obtain the charge ordered state as shown in Fig. 1 (b) (*zigzag model 1*),[13,14] with the unit cell of $2a \times 2b \times 4c$ as shown by a broken rectangle. The unit cell at room temperature is shown by the thin solid rectangle. Because of a disappearance of the mirror symmetry in the $ab$ plane, this model is inconsistent with the space group $Fmm2$. In the second model based on the structure analysis, three kinds of V atoms, $V^{4+}$, $V^{5+}$, and $V^{4.5+}$, are considered; the $16e$ site is occupied by $V^{4.5+}$, and the two $8d$ sites are occupied by $V^{4+}$ and $V^{5+}$. Then, two types of ladders, a charge disordered ladder of $V^{4.5+}$ and the *zigzag* ladder of $V^{4+}$ and $V^{5+}$, coexist below $T_C$ as shown in Fig. 1 (c) (*zigzag model 2*), which is a half of charge ordering, i.e. average $\delta_c \sim 0.5$. The unit cell is again $2a \times 2b \times 4c$ as shown by the broken rectangle. This ordered state is also consistent with recent X-ray structure analysis.[11] Consequently, results of the NMR and the structure analysis lead to two different models, but the existence of a *zigzag* type charge ordered ladder along the $b$-axis is common to both models. The scattering intensity of the $\delta_c$ term has a similar energy dependence in both models, but the intensity and its $Q$-dependence must be different so that the two models can be assessed.

The superlattice intensities are calculated based on the *zigzag model 1, 2* and also on the *chain model* indicated by solid, thin solid, and broken lines in Fig. 3. We used the atomic displacement as $\delta \sim 0.02$ Å for *zigzag model 1*, and the structural parameters determined by Lüdecke *et al.* for *zigzag model 2*. The values of $\delta$ in both models are same order. At $(\frac{15}{2}, \frac{1}{2}, \frac{1}{4})$ the strong intensity modulation resulting from the $\delta_c$ term near $E_A$ can be explained by the calculation of *zigzag model 1, 2*, and is entirely absent in the chain model. As mention above, the shape of modulations is almost independent of the atomic displacement patterns, namely *zigzag model 1, 2*, because the intensity modulation results solely from the $\delta_c$ term. On the other hand, the ratio of the intensity near $E_A$ to that at $E \ll E_A$ is different, which depends on the value of $\delta_c$. As a result, the *zigzag model 1* with $\delta_c$=1 can reproduce the observation better than the *zigzag model 2*. Also at $(\frac{13}{2}, \frac{1}{2}, \frac{1}{4})$, the small intensity modulation around $E \sim 5.5$ keV is reproduced only by the *zigzag model 1*, although the intensity is dominated by the $\delta$ term. We also measured the energy dependence of several other superlattice peaks, which is also consistent with the *zigzag model 1*.

We measured the temperature dependence of the order parameter for this charge ordering. The difference of anomalous scattering factor between $V^{4+}$ and $V^{5+}$ is so small at $E \ll E_A$ that the superlattice intensity is dominated by the $\delta$ term. On the other hand, the intensity of $(\frac{15}{2}, \frac{1}{2}, \frac{1}{4})$ at $E \sim 5.47$ keV results dominantly from the $\delta_c$ term. Therefore the order parameters of atomic displacement ($\delta$) and charge ordering ($\delta_c$) can be separately observed as a function of temperature at $E = 5.41$ ($\ll E_A$) and 5.47 keV($\sim E_A$), respectively. These intensities normalized at 7 K are shown in Fig. 4. Both behave similarly, and are consistent with a continuous second order transition at $T_C$. This is the first direct observation of $\delta_c$. It is clear that charge ordering gradually develops below $T_C$ as the atomic displacement. The present result is incontrast to the NMR data indicating $\delta_c$ varying slightly with temperature.[19] One can see that the slightly different behavior appears near $T_C$. There are two possibilities; One is that the charge order and the atomic displacement take place at different temperatures. The other is that $\delta$- and $\delta_c$-terms may have different critical exponents. It is important to understand in this regard that the charge and lattice are coupled in this system. Such a detailed work is under way.

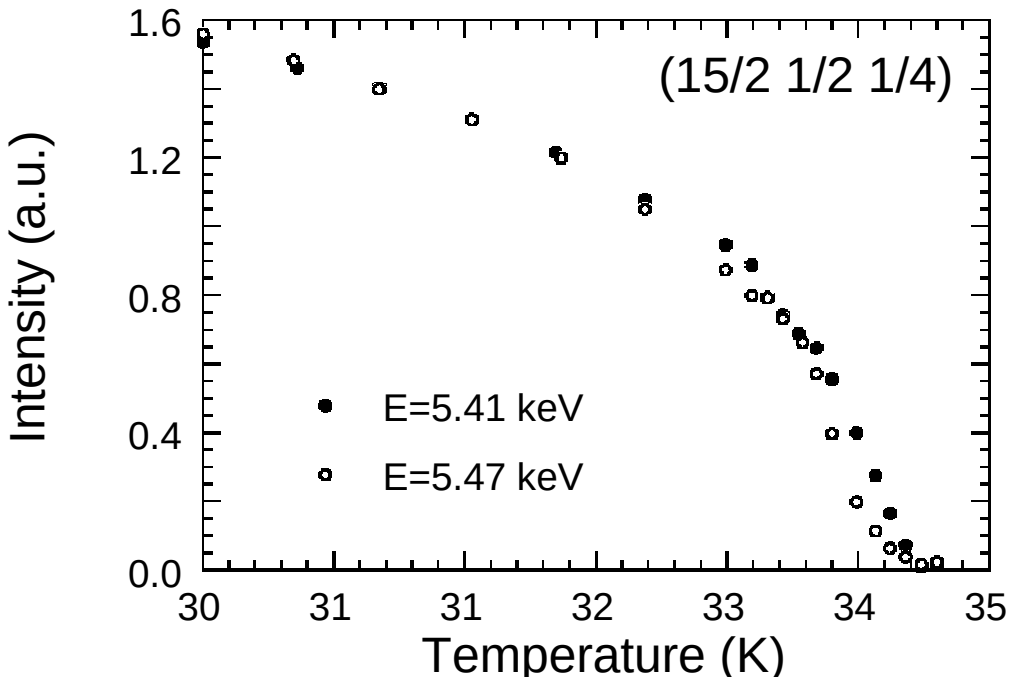

FIG. 4. Temperature dependence of superlattice reflection $(\frac{15}{2}, \frac{1}{2}, \frac{1}{4})$ as measured at $E = 5.41$ keV $\ll E_A$ to probe atomic displacement ($\delta$), and at $E = 5.47$ keV $\sim E_A$ to enhance the charge contribution.

In conclusion, we have reported direct evidence of the fully-charged *zigzag*-type ordered state below $T_C$ by monitoring a dramatic energy dependence of the superlattice intensities at the vanadium K-absorption edge. Charge ordering ($\delta_c$) was observed to gradually develop below $T_C$, manifesting a second order transition.

We would like to thank Sander van Smaalen, Thomas T. M. Palstra, Ken Finkelstein, John M. Tranquada, G. Shirane, and M. Nishi for fruitful discussions. This study was supported in part by the U.S.-Japan Cooperative Program on Neutron Scattering between DOE and MONBUSHO. H.N. acknowledges support by the Japan Society for the Promotion of Science for Young Scientists.

---

* Present address: Photon Factory, Institute of Materials Structure Science, High Energy Accelerator Research Organization(KEK), Tsukuba, 305-0801, Japan.